
\input phyzzx
\unnumberedchapters

\def\third{{1\over 3}}
\def\absh{|H|}
\def\gb{\bar{g}}
\def\nabb{\bar{\nabla}}
\def\ada{{\dot{a}\over a}}

\REF\mp{S.D. Majumdar, Phys. Rev. {\bf 72}, 930 (1947),
A. Papepetrou, Proc. Roy. Irish Acad. {\bf A51}, 191 (1947)}
\REF\romans{L. Romans, Nucl. Phys. {\bf B393}, 395 (1992).}
\REF\gibbons{G.W. Gibbons and S.W. Hawking, Phys. Rev. {\bf D15}, 2738 (1977).}
\REF\hartle{J.B. Hartle and S.W. Hawking, Commun. Math. Phys. {\bf 26},
87 (1972).}
\REF\wilczek{C.F.E. Holzhey and F. Wilczek, Nucl. Phys. {\bf B380}, 447
(1992).}
\REF\kallosh{R. Kallosh, Phys. Lett. {\bf B282}, 80 (1992).}
\REF\ghs{D. Garfinkle, G. Horowitz and A. Strominger, Phys. Rev. D {\bf 43},
3140 (1991).}
\REF\linde{R. Kallosh, A. Linde, T. Ortin, A. Peet and A. Van Proeyen,
preprint SU-ITP-92-13
(1992).}
\REF\wald{R.M. Wald, Ann. Phys. {\bf 82}, 548 (1974).}
\REF\moss{F. Mellor and I. Moss, Phys. Lett. {\bf B222}, 361 (1989).}
\REF\mcvittie{G.C. McVittie, Mon. Not. R. Astron. Soc. {\bf 93}, 325 (1933).}
\REF\eisen{J. Eisenstadt, Phys. Rev. D {\bf 11}, 2021 (1975).}
\REF\inprep{M. Barua, K. Das, D. Goswami, D. Kastor, K.D. Krori
 and J. Traschen, in preparation.}
\REF\gibbonshull{G.W. Gibbons and C.M. Hull, Phys. Lett {\bf 109B}, 190
(1982).}
\REF\jerome{J. Gauntlett, D. Kastor and J. Traschen, work in progress.}

\def\ackn{\foot{This work was supported in part by NSF grant
NSF-THY-8714-684-A01}}
\Pubnum={UMHEP-380}
\date{November, 1992}
\titlepage
\title{Cosmological Multi-Black Hole Solutions\ackn}
\author{David Kastor and Jennie Traschen}
\address{Department of Physics and Astronomy\break
         University of Massachusetts\break
         Amherst, Massachusetts 01003}
\vfil
\abstract
We present simple, analytic solutions to the Einstein-Maxwell equation,
which describe an arbitrary number of charged black holes in a spacetime with
positive cosmological constant $\Lambda$.  In the limit $\Lambda=0$, these
solutions reduce to the well known Majumdar-Papapetrou (MP) solutions.  Like
the MP solutions, each black hole in a $\Lambda >0$ solution has charge $Q$
equal to its mass $M$, up to a possible overall sign.  Unlike the $\Lambda = 0$
limit, however, solutions with $\Lambda >0$ are highly dynamical.  The black
holes move with respect to one another, following natural trajectories in the
background deSitter spacetime.  Black holes moving apart eventually go out of
causal contact.  Black holes on approaching trajectories ultimately merge.
To our knowledge, these solutions give the first analytic description of
coalescing black holes.  Likewise, the thermodynamics of the $\Lambda >0$
solutions is quite interesting. Taken individually, a $|Q|=M$ black hole is in
thermal equilibrium with the background deSitter Hawking radiation.  With more
than one black hole, because the solutions are not static, no global
equilibrium temperature can be defined.  In appropriate limits, however, when
the black holes are either close together or far apart, approximate equilibrium
states are established.

\endpage

\chapter{Introduction}
In this paper we give simple, analytic
solutions to the Einstein-Maxwell equation, which
describe collections of
charged black holes in a spacetime with positive
cosmological constant $\Lambda$.
In the limit of vanishing cosmological constant, our solutions reduce to the
well known Majumdar-Papapetrou (MP) solutions [\mp].
For all values of $\Lambda\ge 0$,
individual black holes in the solutions have charge $Q$ equal to mass $M$,
up to a possible overall sign.  For $\Lambda >0$, however, this is no longer
the
condition for extremality [\romans].
We study the
mechanical and thermodynamic properties of the $\Lambda > 0$ solutions,
which turn out to generalize the
properties of the MP solutions in interesting ways.

For $\Lambda=0$, the black holes are static.
In contrast, the black holes in
a $\Lambda > 0$ solution are highly dynamical.
The black holes ignore one another and
follow natural trajectories in the background deSitter spacetime.
The black holes eventually either merge, or move out of causal
contact.
As far as we know, these solutions give the first analytic
description of coalescing black holes.  However, there are many
questions about the causal structure and dynamics of these solutions
which remain to be answered.

The thermodynamics of solutions with $\Lambda > 0$ is also
quite interesting.  The black holes in the MP solution each have
vanishing Hawking temperature, and so are in thermal equilibrium,
with each other and also with the background flat spacetime.
With $\Lambda >0$, the background deSitter
spacetime has an ambient Hawking temperature [\gibbons].
A single $|Q|=M$ black hole actually has a temperature
equal to the deSitter temperature, and hence would be in a
state of thermal equilibrium with the background spacetime.
The temperature, however, depends on both the cosmological constant and the
mass of the black hole.  Adding more black holes, with differing masses,
then does not produce a thermal equilibrium state.
Indeed, with more than one black hole, because
the solutions are not static, no global
equilibrium temperature can be defined.
Approximate temperatures, however,
can be defined in the limits where the black holes
are either widely separated or coalesced.
A given solution, then, describes a transition between one
state of thermal equilibrium in the far past and another equilibrium state in
the far future.  However, an understanding of the
nonequilibrium thermodynamics of the transition between these states will
require further investigation.

\chapter{Majumdar-Papapetrou Solutions}
In Newtonian mechanics, a collection of charged point particles, each having
charge equal to its mass\foot{We use geometrical units, $G=c=\hbar=1$},
can stay at rest in a state of mechanical equilibrium.
The electrostatic repulsions of the particles exactly balance the
gravitational attractions.
Remarkably, the same balance holds in general relativity.
The MP solutions to the source free Einstein-Maxwell
equation correspond to this Newtonian situation.  The MP solutions,
themselves, are geodesically incomplete.  Hartle and Hawking [\hartle] showed
how the MP solutions could be analytically extended and interpreted as a system
of charged black holes.
The metric and gauge field for the MP solutions are given by
$$\eqalign{ds^2=-\Omega^{-2}dt^2 +\Omega^2 \left(dx^2+dy^2+dz^2\right ),\qquad
A_t= \Omega^{-1},&\cr  \Omega=1+\sum_i {m_i\over r_i},\qquad
r_i=\sqrt{(x-x_i)^2+(y-y_i)^2+(z-z_i)^2},&}\eqn\mpsoln$$
where $m_i$ and $(x_i,y_i,z_i)$ are the masses and locations of the black
holes.
It can be shown [\hartle]
that the points $r_i = 0$ actually represent event horizons
of area $4\pi m_i^2$.  For the case of one black hole, the metric
\mpsoln\ is just the extreme Reissner-Nordstrom metric in isotropic
coordinates.

A single Reissner-Nordstrom black hole having charge equal to its mass
is the simplest example of an extremal black hole.  If the charge of the hole
were increased (or the mass decreased) further, then the curvature singularity
would no longer be hidden behind an event horizon.  If such a naked singularity
arose in a `physical' solution, it would violate the cosmic censorship
conjectures and lead to a breakdown of predictability.
That the Hawking temperature of an extremal
Reissner-Nordstrom black hole vanishes is impressive evidence for
cosmic censorship.  The evaporation of a charged black hole
terminates when it reaches the extremal state, leaving the singularity
hidden\foot{Holzhey and Wilczek
[\wilczek] have studied how the evaporation process for charged dilaton
black holes can terminate at an extremal state, even though the temperature
of this state may be nonzero}.  The MP solutions then describe
charged black holes in thermal, as well as mechanical, equilibrium with
temperature equal to zero.
A related property is that the MP solutions
are exact ground states of $N=2$ supergravity.  Since all quantum
corrections to the effective action expanded around the MP solutions
vanish [\kallosh].
Analogues of the MP solutions have also
been written down for dilaton black holes [\ghs,\linde].
The individual black holes in these solutions are also extremal ones.

\chapter{Reissner-Nordstrom-deSitter Solutions}
There is an analogue of the Reissner-Nordstrom solution for a spacetime
with a cosmological constant.  The Reissner-Nordstrom-deSitter (RNdS) metric
and gauge field in Schwarzschild coordinates are given by
$$\eqalign{ds^2=-V(R)&dT^2 + V^{-1}(R)dR^2 +R^2d\Omega^2,\qquad
A_T=-{Q\over R},\cr &
V(R)=1-{2M\over R}+{Q^2\over R^2}-{1\over 3}\Lambda R^2,}\eqn\metric$$
where $M$ and $Q$ are the mass and charge of the hole and $\Lambda$ is the
cosmological constant.
The metric has a curvature singularity
at the origin.  There are event horizons at the radii where
$V(R)$ vanishes.  Taking $M=Q=0$ in \metric\ gives the static form of the
deSitter metric.

Romans [\romans] has recently studied the
thermodynamics of these solutions.  This is more complicated, because
for $\Lambda > 0$ the
deSitter Horizon also radiates at its own temperature [\gibbons].
Indeed, Gibbons and Hawking [\gibbons] have extended the laws of black
hole thermodynamics to include cosmological event horizons.
The Hawking temperature for a horizon at radius $\rho$ is given by
$$T={|\kappa |\over 2\pi}={1\over 4\pi}
\left |V^\prime(\rho)\right |, \eqn\temp$$
where $\kappa$ is the surface gravity at the horizon.
As in the $\Lambda=0$ case,
Romans finds extreme RNdS black holes with
zero temperature, in which the inner and outer black hole horizons coincide.
For $\Lambda > 0$, the extremal black holes have $M< |Q|$.
In Appendix A, we show that, as in the
$\Lambda=0$ case [\wald], it is
impossible to destroy the horizon of an RNdS black hole by
throwing in charged particles to charge it past the extremal limit.

Another interesting class of RNdS solutions have the
temperatures of the black hole and deSitter horizons equal.
Remarkably,
these solutions have $|Q|=M$\foot{This had been noted previously by
Mellor and Moss [\moss]}.
The metric function then
has the form
$$ V(R)= (1-{M\over R})^2-\third \Lambda R^2 \eqn\metricfunction$$
The common temperature is given by [\romans]
$$T={1\over 2\pi}
\sqrt{{\Lambda\over 3}\left ( 1-4M\sqrt{\Lambda /3}\right )}.
\eqn\temperature $$
In the naive picture of black hole evaporation, the $|Q|=M$
solutions are thermodynamically stable states and are the end points of the
evaporation process.
If $M > |Q|$ then the black hole
is hotter than the deSitter horizon and it will evaporate until it reaches
$M=|Q|$.
If $M<|Q|$, then
the deSitter horizon is hotter and the black hole will accrete radiation until
it reaches $M=|Q|$.  It is interesting that for $\Lambda >0$ the conditions of
extremality and thermal equilibrium no longer coincide.

We note that the total gravitational entropy
of an RNdS black hole is maximized,
for fixed charge and cosmological constant,
in this equilibrium state.  The total gravitational entropy
is given by the sum of the areas of the black hole and cosmological event
horizons,
$$S_{grav}={1\over 4}A_{BH}+{1\over 4}A_{dS}. \eqn\entropy$$
The result then follows from the generalized first law of thermodynamics
given in [\gibbons].  For an infinitesimal perturbation between RNdS solutions
of fixed charge, the generalized first law states that
$$\kappa_{dS}\delta A_{dS} -\kappa_{BH}\delta A_{BH}=0,\eqn\firstlaw$$
where $\kappa_{dS}$ and $\kappa_{BH}$ are the surface gravities of the deSitter
and black hole event horizons.
The change in entropy of such a variation can then be written as
$$\delta S_{grav}={1\over 4}\left(1+{\kappa_{BH}\over\kappa_{dS}}\right)
\delta A_{BH}. \eqn\deltaentropy$$
Clearly, the entropy is extremized for $\kappa_{dS}= -\kappa_{BH}$,
which coincides with the condition that the black hole and deSitter
temperatures be equal, and happens
when $|Q|=M$.  From the discussion in the previous paragraph, we see that the
extremum is a maximum.
This result is in contrast to an evaporating Schwarschild black hole, where
the gravitational entropy decreases.  Here, the deSitter horizon acts like the
walls of a box.  The entropy of the
deSitter horizon measures, in some sense, the entropy of the Hawking radiation
carried away from, or absorbed by the black hole.

\chapter{Motion of Test Particles}
Are cosmological analogues of the MP solutions built out of extremal
or $|Q|=M$ black holes? The black holes in the MP solution ``ignore''
one another.
We would like to find some similar phenomenon for charged black holes
in a deSitter background.
To find the right criterion, we look at
the motion of charged test particles in the RNdS metric \metric .
We will see that the motion of $q=m$ test particles in a $Q=M$
RNdS background is particularly simple.

The conserved energy $E$ of a test
particle of charge $q$, mass $m$ and $4$-velocity $u^a$ on a radial
path in an RNdS background is
$$\eqalign{{E\over m} &= -\xi^a\left(u_a+{q\over m}A_a\right) \cr
& = V(R){dT\over d\tau} -{q\over m}A_T(R),}\eqn\energy$$
where $\xi^a$ is the static Killing vector and $\tau$ is the proper time.
Together with the normalization condition $u^a u_a = -1$,
this gives the equation of motion
$$\left ({dR\over d\tau}\right)^2 = - V(R) +
({E\over m} + {q\over m} A_T (R))^2. \eqn\eqofmot$$
Substituting in the gauge field and metric function for a
$Q=M$ black hole, this becomes
$$\left ({dR\over d\tau}\right)^2= - \left( 1- {M\over R}\right ) ^2
+\third \Lambda R^2 + \left(1-{qM\over mR}\right)^2.  \eqn\eomot  $$
If the test particle has $q=m$ and has
energy to equal its rest mass (i.e. $E=m$), then this further
reduces to
$${dR\over d\tau} =\pm\sqrt{{\Lambda\over 3}}R. \eqn\eomota $$
This in turn is the same as the equation of motion for a minimum energy
test particle in a background deSitter spacetime (i.e. $Q=M=0$).
This looks like what we want.  The $q=m$
test particle is, in some sense, not affected by the presence of the
$Q=M$ black hole.   This hint will turn out to be what we need to guess an
exact solution.

Note that there are two choices for the path of the $q=m$ test particle
in \eomota .  It can be either ``ingoing'' or ``outgoing''.  This is in
contrast
to the $\Lambda = 0$ case, where a
minimum energy $q=m$ test particle stays at rest in
the field of a $Q=M$ black hole.  Choosing one path or another breaks the
time reversal invariance of the system.  We will continue to use the names
``ingoing'' and ``outgoing'' to denote these paths below, even though in
different coordinates they may not look ingoing or outgoing.


\chapter{Cosmological Coordinates}
We seek to promote our $q=m$ test particles to black holes,
expecting that they will follow
``ingoing'' or ``outgoing'' paths as in \eomota .
Such black hole solutions should be quite complicated in static coordinates.
For example, moving charged black holes will generate magnetic as
well as electric fields.  However, in cosmological coordinates the motion of
such minimum energy particles is quite simple.

The RNdS solutions can be rewritten in cosmological coordinates.
For $Q=M$ this has the form
$$\eqalign{&ds^2=-\Omega^{-2}dt^2 +
a^2(t)\Omega^2\left(dr^2+r^2 d\Omega^2\right),
\qquad A_t=\Omega^{-1}, \cr
 & \Omega = \left (1+{m\over ar}\right ),\qquad a(t)=e^{Ht},\qquad
H=\pm\sqrt{{\Lambda\over 3}}  } \eqn\cosmornds$$
For $M=0$ this is just the standard cosmological form of the deSitter
metric.
We will call the coordinate system with $H>0$ expanding and that
with $H<0$ contracting.
The static Killing vector is given by
$$\xi^a = \left ({\partial \over\partial t}\right)^a -
 H r\left ({\partial\over\partial r}\right )^a. \eqn\killingvector $$

The norm of the
Killing vector vanishes at the horizons, which implies that the
deSitter horizon $r_{+}$
and the outer black hole horizon $r_{-}$ are located at
$Har_\pm\Omega^2=1$, or
$$r_\pm = {1\over 2a(t)\absh}\left ( 1-2M\absh\pm\sqrt{1-4M\absh}\right ).
\eqn\horizon$$
There is also an inner black hole horizon at negative $r$ in these coordinates.
Note that the products $a(t)r_\pm$ are constants and also that the metric
\cosmornds\ is non-singular at $r_{\pm}$.
One can check that the surface gravity $\kappa$ at the two horizons,
defined by
$$\half\nabla_a \left (\xi^b\xi_b \right )=-\kappa\xi_a,\eqn\surfg $$
yields the correct value \temperature\ for their common temperature.

The transformation between the static and cosmological coordinates is
given by
$$ a(t)r=R-M, \qquad t=T+h(R), \qquad
h^\prime(R)= - {H R^2\over (R-M)V(R)} \eqn\coordtrans $$
Integrating this transformation near the black hole and deSitter horizons,
one finds that the expanding coordinates cover the past black hole horizon
and the future deSitter horizon.  Likewise, the contracting coordinates cover
the past deSitter horizon and the future black hole horizon.  Thus, the metric
\cosmornds\ with $H>0$ actually describes the white hole portion of the
extended
spacetime, while the metric with $H<0$ describe the black hole part.
This can be confirmed by looking at null geodesics.  The 2-sphere $r=r_{-}$
is an outer trapped surface for $H<0$.  Null rays cannot get out.  For $H>0$
it is an inner trapped surface.  Null rays cannot get in.

We can look at the paths of $q=m$ test particles in \cosmornds .  The
paths \eomota\ that were ``outgoing'' in static coordinates, stay at constant
spatial comoving coordinate in the expanding coordinates, whereas they move out
``doubly" fast in the contracting coordinates (and vice-versa
for the ``ingoing'' test particles).

It is useful in understanding the new solutions below to look at the motion of
these particles in a little more detail.  Consider an ``outgoing" $q=m$
test particle in expanding coordinates ($H>0$).
It stays at a constant coordinate
radius.  The deSitter and white hole horizons $r_\pm$, however, redshift like
$1/a$.
At early times the horizons are both at large radii and the test
particle is inside both.  At some later time the white
hole horizon sweeps past it, and then at some still later time
the deSitter horizon sweeps past it.
In static coordinates, we could follow the portion of the particles
path between the two horizons.  We would see it leave the white hole horizon at
$T=-\infty$ and move out to the deSitter horizon at $T=+\infty$.  Likewise,
an ``incoming'' test particle, which stays at a fixed radial coordinate in
contracting coordinates ($H<0$), starts outside the past
deSitter horizon, moves in
through it and then through the black hole horizon.  In static coordinates,
this looks like a particle leaving the deSitter horizon at $T=-\infty$ and
getting to the black hole horizon at $T=+\infty$

\chapter{Cosmological MP Solutions}
The metric \cosmornds\ for one $|Q|=M$ RNdS black hole in cosmological
coordinates closely resembles the MP metric \mpsoln\ for the case of one black
hole.  This suggests a simple form for the cosmological MP solutions
$$\eqalign{
ds^2=-\Omega^{-2}dt^2 + a^2(t)\Omega^2 \left(dx^2+dy^2+dz^2\right ),\qquad
A_t= \Omega^{-1},\qquad a(t)=e^{Ht},&\cr
\Omega=1+\sum_i {m_i\over ar_i},\qquad
r_i=\sqrt{(x-x_i)^2+(y-y_i)^2+(z-z_i)^2},\qquad
H=\pm\sqrt{{\Lambda\over3}} .&}\eqn\cosmomp$$
Indeed, we will show that more generally, a metric and gauge field
of the form \cosmomp\ solve the Einstein-Maxwell equations with cosmological
constant $\Lambda$, if $\partial_t\left( a(t)\Omega\right)=\dot{a}$ and
$\nabb^2\Omega = 0$, where $\nabb^2$ is the flat space Laplacian.

First consider the constraint equations.
The Hamiltonian, momentum and Gauss' law
constraints on a spatial slice are given by
$$\eqalign{&H=-{}^{(3)}R + {1\over g}\left (\pi^{ij}\pi_{ij}-\half\pi^2\right)
=-16\pi\rho \cr
&H_k=-{2\over\sqrt{g}} {}^{(3)}\nabla_i\pi^i_k=16\pi J_k,\cr
&{}^{(3)}\nabla_iE^i=0, }\eqn\constraints $$
where $g_{ij}$ and $\pi_{ij}$ are the spatial metric and momentum,
$\rho$ and $J_k$ are the energy and current densities, and $E^i$ is the
electric field.
For a metric of the form \cosmomp ,
if we assume that $\partial_t(a\Omega)=\dot{a}$ then the momentum is given by
$$\pi_{ij}=-2\ada\sqrt{g}g_{ij}.  \eqn\rwmom$$
Given this simple relation between $\pi_{ij}$ and the metric, the  momentum
constraint in \constraints\ is satisfied with zero current.
The $3$-dimensional scalar curvature is
$${}^{(3)}R={1\over a^2\Omega^2}\left\{-{4\over\Omega}\nabb^2\Omega
+{2\over\Omega^2}\gb^{ij}(\nabb_i\Omega)\nabb_j\Omega\right\},\eqn\curvature$$
where $\gb_{ij}$, $\nabb_i$ are the flat spatial metric and derivative.
The energy density $\rho$ has contributions from the Maxwell field and the
cosmological fluid $\rho=\rho_{max}+\rho_{cos}$.  If $A_t=1/\Omega$, then
$$8\pi\rho_{max}={1\over a^2\Omega^4}\gb^{ij}(\nabb_i\Omega)
\nabb_j\Omega.\eqn\energydensity $$
{}From the expressions \rwmom , \curvature\ and
\energydensity , one sees that the Hamiltonian
constraint can be satisfied by having the ``black hole parts'' and
the ``cosmological parts'' vanish separately.  That is, one has a solution if
$$\nabb^2\Omega=0,\quad{\rm and}\quad \left(\ada\right)^2= {8\pi\over 3}
\rho _{cos}.\eqn\conditions$$
Further, the Gauss' law constraint is satisfied if \conditions\ holds.
Note that \conditions\ can be satisfied by any time dependent, but spatially
constant $\rho_{cos}$.

Some insight into the nature of the solutions
is gained from the matter equations of motion, $\nabla _a (T^{ab}_{max}
+T^{ab}_{cos})=0$. The time component of this equation is
$$ {1\over {\Omega}}{d\over {dt}} \rho _{cos} +3\ada (p_{cos} +
\rho _{cos})+{1\over 2\Omega}{d\over dt}\rho_{max}+
2\ada \rho_{max}=0\eqn\timedependence$$
{}From \energydensity\ it follows that the Maxwell terms
themselves sum to zero.  Therefore, the part involving the cosmological fluid
must also vanish independently. At this point there are two choices for
a solution.   From \conditions , $\rho_{cos}$ cannot depend on the spatial
coordinates.
If $\rho _{cos}$ is allowed to be time dependent,
then, since $\Omega$ has spatial dependence, the
pressure $p_{cos}$ must be spatially dependent.
This makes sense physically; ordinary matter would tend to accrete arround
the black hole.  The inhomogeneous pressure is needed to keep the density
constant.  In the one black hole case, such solutions have been studied by
McVittie [\mcvittie] and others (see e.g. [\eisen]).
The other choice, which
we shall make, is to take $\rho_{cos}$ to be independent of time.
We then have $p_{cos} =-\rho _{cos}=-\Lambda $, the form for a cosmological
constant. Again,
this makes sense physically,
because a cosmological constant cannot accrete\foot{An idealized
model of a cosmic string, spacetime minus a wedge, can be embedded
in any flat Robertson-Walker background since the string does not accrete
[\inprep]}.
Finally, we note that the full set of evolution equations
for $\pi_{ij}$ are straightforward to check and yield no more constraints.

\chapter{Geometric and Thermodynamic Properties}
The solutions \cosmomp\
with $H<0$ appear to describe a system of ``incoming'' charged black holes.
The solutions with $H>0$ would
describe the time reversed situation; a system of ``outgoing'' charged white
holes.
The first thing to establish is that the solutions really do contain black hole
(or white hole) horizons.  One expects this to be the case.  However, it is not
straightforward to locate the horizons.  The solutions, with more than
one black hole, do not appear to have a stationary Killing vector.  Thus,
one cannot
simply look for the surfaces on which the Killing vector becomes null.  One can
look for apparent horizons (boundaries of regions of trapped surfaces)
in a given spatial slice, but this too is
complicated by the lack of symmetry.

The situation does simplify for early and late times, when
the holes are either far apart or close together.
For concreteness, consider two ``incoming''
black holes in a background with $H<0$.
At early times the holes are far apart, and near each one the metric approaches
that for a single hole.
In this limit, one can verify that there are regions of
outer trapped surfaces surrounding each of the holes.
These regions extend out to radii $r_{-,i}$
given by \horizon .
As time goes on, the universe contracts, and
the coordinate size of each of these apparent horizons grows.  Their shapes
will distort due to the presence of the other hole.  At some point in time
the two apparent horizons will collide with one another and presumably merge.
Indeed, in the late time limit one can verify that outer trapped surfaces
surround both of the holes.

To summarize, when $H<0$, one can show that the two objects are first
surrounded individually by outer trapped surface, and then later there is an
outer trapped surface that surrounds both.  For white holes with $H>0$,
the situation is time reversed.  At early times
inner trapped surfaces surround both objects together.
Later on, the objects are surrounded only separately by regions of inner
trapped surfaces.
These results agree with what one expects from the area theorem.  Black holes
merge, and white holes split.

Another question is whether an extension of the solutions with $H>0$,
which covers the black
hole portions of the ``outgoing'' holes (and likewise for
the white hole portions
of the ``ingoing'' black holes with $H<0$).
For the case of one black hole, \coordtrans\ gives
an explicit coordinate transformation
from expanding to contracting coordinates.  It appears that making this
transformation locally about one of the ``outgoing'' holes
(as $t\rightarrow\infty$),
does indeed extend the spacetime to cover a black hole horizon.  However,
aspects of this transformation are still confusing; e.g. how the different
regions fit together.  We defer an explicit presentation
to future work.

For the purposes of discussion in this paper,
we will assume that the ``outgoing'' white hole spacetimes
are extendable to coordinates that cover black hole horizons and past
deSitter horizons.  Likewise, the ``incoming''
black hole solutions are assumed to be fully extendable.
A related question is whether
the black holes must all be ``outgoing'' or all ``incoming'', or whether
there can be arbitrary combinations of ``incoming'' and ``outgoing'' holes.

Although a full understanding of the thermodynamics of these solutions must
await a better understanding of their analytic extensions and horizon
structures, the thermodynamics appears to be quite interesting.
Given that there is no stationary Killing vector, the usual
definition of temperature in terms of surface gravity at a horizon \temp\
does not work.  Indeed, these solutions appear to be highly non-equilibrium.
Still,
as for the horizon structure, we can give a simple description of the
thermodynamic behavior
at early and late times, as follows.

Recall that for one $Q=M$ RNdS black hole, the deSitter and black hole horizons
are in thermal equilibrium at a common temperature $T_{bh}=T_{dS}=T(M,\Lambda)$
given by \temperature .
Now consider two black holes in a deSitter background,
both satisfying $Q=M$, but with different masses $M_1$ and $M_2$.
If the black holes are widely enough separated
to be out of causal contact, then
each black hole will have its own distinct deSitter horizon.
Also, in the region near each black hole there will be
an approximate static Killing vector that can be used to define an approximate
temperature.  Each of the black holes will be in
approximate thermal
equilibrium with its own deSitter horizon at temperatures $T(M_1,\Lambda)$
and $T(M_2,\Lambda)$ respectively.
If the two black holes are ``ingoing'' with respect to one another, then
this will be the situation at very early times.  Later the black holes
come into causal contact and eventually merge into a single black hole with
mass $M=M_1+M_2$.  At very late times, there will again be an approximate
static Killing vector.  The black hole horizon will be in equilibrium with
the deSitter horizon at temperature $T(M_1+M_2,\Lambda)$.

\chapter{Conclusions and Further Questions}
We have presented solutions to the Einstein-Maxwell equations with a
positive cosmological constant, which plausibly represent a collection of
charged black holes moving either towards or away from one another.
We have also given a description of the horizon structure and thermodynamics
of these solutions in the early and late time limits.  There is clearly a
good deal more work to be done on these solutions; some of which has been noted
in the previous section.

Let's start with questions about the classical properties of the solutions.
It is important to know whether these solutions can be fully
analytically extended, as has been assumed above, and to have a clear picture
of their horizon structures.  It should be especially interesting
to study the regime in which the black holes are merging (the white holes
splitting).  These solutions might give insight into an approximation which
treats astrophysically interesting mergers of black hole binaries.  Although,
note that here the mergers take place without any gravitational radiation.

The thermodynamics of the intermediate state, where the black holes are
distinct, but still in causal contact, should be interesting.
Will the masses of the black holes change due to emission and absorption
of Hawking radiation during this period?  If so, then they should
emerge from this nonequilibrium state with charges in general differing from
their masses.  Eventually then, each black hole should exchange radiation
with the background until it again reaches a $Q=M$ state.
A first step towards understanding the
exchange of energy, would be to study what a particle detector sees, if
follows the path of a $q=m$ test particle in a $Q=M$ RNdS background.

The splitting and merging of holes raises interesting questions about the
parameter space of
$\Lambda=0$ solutions.  Extreme RN black holes are regarded as solitons of
general relativity, satisfying a kind of Bogomolnyi bound [\gibbonshull].
But in at least one respect they appear to differ from other solitons.
Consider a magnetic monopole in the Bogomolnyi limit with two units of
topological charge.  The solution will have zero modes corresponding to
the possibility of breaking it up into two monopoles, each having unit
topological charge.  By analogy, we would expect an extreme RN black hole
to have zero modes, corresponding to the possibility of
breaking it up into smaller extreme black
holes.  By the area theorem, though, this cannot be the case.
On the other hand, a single $Q=M$ RNdS white hole appears to be unstable.
It can be split into any number of charge equal mass fragments,
which are then carried apart by the expansion of the universe.  In the limit
of zero cosmological constant, this may correspond to marginal stability
of an extreme RN white hole.  This picture is supported by analysis at the
level of test particles.  A $q=m$ test particle can stay at rest in a $Q=M$ RN
background, whether it is inside or outside the event horizon.  Hence, there
should be analogues of the MP solutions describing merged black holes.

Another interesting set of questions involves supersymmetry.
Multi-object solutions are usually associated with Bogomolnyi bounds arising
from an
underlying supersymmetry of the solution.  Romans [\romans] has noted, though,
that the relevant supersymmetry (coming from $N=2$ Yang-Mills supergravity)
is consistent only with $\Lambda<0$.  Our
solutions, then, are not supersymmetric, at least in this sense.
On the other hand, for
$\Lambda<0$, while the $Q=M$ RNdS holes are supersymmetric, they
are also naked singularities.
It should be interesting to see whether these can also be assembled into
multi-hole solutions and to understand the role played by supersymmetry
[\jerome].

\Appendix{A}
In reference [\wald], Wald asked whether you can destroy a black hole
by overcharging it.  One might think that, by throwing in particles with
a high charge to mass ratio, one could charge the hole past the extremal limit
of $Q=M$.  It turns out [\wald] that for a charged particle to get over
the Coulomb barrier into the hole, it must have more energy greater than
or equal to its charge.

Here we do the analogous calculations for the extreme RNdS black holes
[\romans].
For a metric of the form
$$ds^2= - V(r)dt^2 +V(r)^{-1}dr^2+r^2d\Omega^2  \eqn\spherical$$
with gauge field $A_t(r)=-Q/r$,
the equation of motion for a test particle (of energy $E$, rest mass $m$ and
charge $q$) on a radial geodesic is given by \eomot
$$ \left ({dr\over d\tau}\right)^2 = - V(r) + {1\over m^2}
\left ( E + qA_t(r) \right )^2. \eqn\eomagain$$
In order for the particle to get into the hole it must, at least, reach the
event horizon.  Looking for the minimum energy such particle, we set $E=m$
and $dr/d\tau = 0$ at the horizon.  At the horizon radius $\rho$, we then have
($V(\rho)=0$)
$$\left (1-{qQ\over m \rho}\right )^2\geq 0.\eqn\bound$$
This translates to
$$ {m\over q}\geq {Q\over \rho}.\eqn\minenergy$$
For the extreme RNdS black holes we have [\romans],
$$ M=\rho(1-{2\over 3}\Lambda \rho^2), \qquad
   Q^2=\rho^2(1-\Lambda \rho^2). \eqn\extreme $$
{}From these we can compute how the mass and charge of an extremal hole
change with each other.  We find
$${\partial M\over\partial Q}= \sqrt{1-\Lambda\rho^2} =
{Q\over \rho},\eqn\dmdq $$
which coincides with the bound \minenergy . Hence an extreme RNdS black
hole cannot be pushed over the limit.  Given the nontrivial relation
\extreme\ between the charge and mass of the extreme black holes, this is
a somewhat striking confirmation of cosmic censorship.

\refout
\end